\documentclass[12pt,preprint]{aastex}

\usepackage{color}

\newcommand{\beq}{\begin{equation}}
\newcommand{\eeq}{\end{equation}}
\newcommand{\beqa}{\begin{eqnarray}}
\newcommand{\eeqa}{\end{eqnarray}}

\title{Primordial Helium Abundance: A Reanalysis of the Izotov-Thuan 
Spectroscopic Sample} 
\author{Masataka Fukugita$^{(a,b)}$ and Masahiro Kawasaki$^{(a)}$}
\affil{$^{(a)}$Institute for for Cosmic Ray Research, University of Tokyo,\\
             Kashiwa 277-8582, Japan}
\affil{$^{(b)}$Institute for Advanced Study, Princeton, NJ 08540 USA}
\begin{document}
\maketitle
\section*{Abstract}
A reanalysis is made for the helium abundance determination for the
Izotov-Thuan (2004) spectroscopic sample of extragalactic H II regions. 
We find that the effect of underlying stellar absorption of the He I lines,
which is more important for metal poor systems, affects
significantly the inferred primordial helium abundance $Y_p$
obtained in the zero metallicity limit and the slope of
linear extrapolation, $dY/dZ$. This brings $Y_p$
 from $0.234\pm0.004$ to $0.250\pm 0.004$ and $dY/dZ=4.7\pm 1.0$ to
$1.1\pm 1.4$.  Conservatively, this indicates the importance of the
proper understanding
of underlying stellar absorption for accurate determinations
of the primordial helium 
abundance to the error of $\delta Y_p\simeq 0.002-0.004$.


\section{Introduction}
Izotov and Thuan (2004; hereinafter IT04) presented the primordial 
helium abundance 
$Y_p=0.242\pm0.002$, consistently with their earlier publications
(Izotov and Thuan 1998, and references therein),
 from helium recombination lines in metal poor extragalactic HII
regions. This is significantly higher than the earlier
values given by a number of authors (Pagel et al. 1992;
Olive et al. 1997;
Peimbert et al. 2000), yet is significantly 
lower by three standard deviations
than the expectation from
the baryon abundance constrained from CMB temperature anisotropies (Spergel
et al. 2003) with the aid of Big-Bang nucleosynthesis \footnote{We adopt 
here the neutron lifetime from the
conventional decay counting of ultra cold neutrons (Eidelman et al.
2004).}.  
Particularly intriguing is the small errors which are common
to nearly all analyses. We suspect that this might not represent
properly the error including systematics.

IT04 provided high quality spectroscopic data for 33 HII regions
detailed enough for us 
to repeat the analysis. In this paper we are particularly concerned 
with the effect induced by underlying stellar
absorption on nebula helium emission lines that is poorly constrained.
Our analysis is along the line of Peimbert et al. (2002) and 
Olive and Skillman (2004; hereinafter OS04),
who introduced the helium absorption strength as a free parameter,
although ours are somewhat more conservative. OS04 and  
Peimbert et al. (2000; 2002) 
presented an analysis in which plasma temperature is also
determined by the helium emission lines alone. However, this, 
in principle proper approach given accurate data, induces 
large errors in the resulting helium abundance with the
present accuracy of the available data for distant HII regions, 
and a trend that
might be brought about with the inclusion of stellar absorption
is buried in the noise. Since we do not see a compelling reason
that plasma temperatures from helium and oxygen are significantly
different, 
we take the approach that plasma 
temperature is determined by the ratio of oxygen emission lines,
as was done in most of the work including IT04, and study the
effect of stellar absorption by introducing it as 
a free parameter. We show that $Y_p$ and $dY/dZ$
are very sensitive to the introduction of the underlying
stellar absorption of the helium lines.

\section{Data and procedures of the analysis}

We consider 30 of the 33 H II regions given in IT04,\footnote{%
We do not include the sample of Izotov \& Thuan (1998) because
the equivalent widths are not provided for the recombination lines.}
discarding 3 H II regions
(UM133, Mrk1063, HS0111+2115), for which He I $\lambda$4026 line is not
detected. We also add NGC346 (Region A) studied by Peimbert et al. (2000).
The 31 H II regions we studied are shown in Table 1, where our
final results are also presented. The plasma temperature of OIII is
determined from the ratio of [O III] emissions $\lambda$4363 versus
$\lambda\lambda$4959,5007. The extinction
and stellar absorption of H I Balmer lines are derived from
H$\alpha$, H$\beta$, H$\gamma$ and H$\delta$ using the recombination
calculation of Hummer \& Storey (1987) for the intrinsic
line intensity ratios $I(\lambda)/I({\rm H}\beta)$, as
\begin{equation}
\frac{I({\rm H}\lambda)}{I({\rm H}\beta)}=
\frac{F(\lambda)}{F({\rm H}\beta)}
\frac{W({\rm H}\lambda)+a_{\rm HI}}{W({\rm H}\beta)+a_{\rm HI}} 
\frac{W({\rm H}\beta)}{W({\rm H}\lambda)}
10^{f(\lambda)c_{{\rm H}\beta}}.
\label{eq:Hemis}
\end{equation}
where $F(\lambda)$ is the observed line intensity, $W(\lambda)$ is the
equivalent width, $a_{\rm HI}$ is the stellar absorption and 
$f(\lambda)c_{{\rm H}\beta}$ is the extinction relative to H$\beta$.
The inclusion of H9 and higher
Balmer lines, where available, does not modify the result beyond our interest. 
H8 is blended with He I$\lambda$3889, and equation (\ref{eq:Hemis}) is
used to deblend the helium line. We discard H7 that is deblended with
[Ne III].
We use the extinction curve of O'Donnel (1994), but the use of
the different extinction curve leads only to
a small difference that can be ignored here.

We consider six He I lines, five lines for orthohelium 
$\lambda$3890 ($3s\rightarrow 2p$),
$\lambda$4026 ($5d\rightarrow 2p$), 
$\lambda$4471 ($4d\rightarrow 2p$), $\lambda$5875 ($3d\rightarrow 2p$), 
$\lambda$7065 ($3s\rightarrow 2p$),
and one line for parahelium 
$\lambda$6678 ($3d\rightarrow 2p$). 
The $\lambda$3890 and $\lambda$7065 are sensitive to a fluorescent 
correction and thus to the radiative transfer. 
$\lambda$4026 is generally  weak, susceptible largely to
stellar absorption. 
We use the effective
recombination coefficients of Benjamin et al. (1999), which include
collisional excitation, and the radiative transfer calculation of
Benjamin et al. (2002) for fluorescent corrections, $f_\lambda$,
which is controlled by the optical depth $\tau(3890)$. 
We calculate
the abundance of singly ionised helium as
\begin{equation}
y^+=\frac{F({\rm HeI}\lambda)}{F({\rm H}\beta)} 
\frac{I({\rm H}\beta)} {I({\rm HeI}\lambda)}
\frac{W({\rm H}\beta)}{W({\rm H}\beta)+a_{\rm HI}}
\frac{W({\rm HeI}\lambda)+a_{\rm HeI}}{W({\rm HeI}\lambda)}
10^{f(\lambda)c_{{\rm H}\beta}}\frac{1}{f_\lambda}.
\end{equation}
This is the same as that adopted by OS04 and that by IT04 up to
the inclusion of $a_{\rm HeI}$. We do not know what ratios
are to be taken for $a_{\rm HeI}$ for different lines.
We assume here that all absorption
strengths are identical in equivalent strengths, as was done in OS04.
This is probably not too bad an approximation in view
of the observation for absorption in B stars (Lennon et al. 1993;
Lyubimkov et al. 2000)
and the  
size of the resulting errors in our calculation
that amount to $\approx$50\% of the
central values: more detailed line ratios are the matter at a
higher order level. 
We refer to Olive \& Skillman (2001) for more detailed
discussion for this issue.

Our method of analysis differs from the usual one to find the parameters.
We find the best likelihood solution in full four parameter space
$y^+$, $n_e$ (electron density), $a_{\rm HeI}$ and $\tau$ for
six lines, rather than minimising the sum of $\chi^2$ for each
line, which has been adopted in the literature, for the purpose to
take the correlation among the lines into account. The electron
density is poorly constrained, but its indeterminacy 
affects the resulting $Y$ little due to its very weak dependence
in the recombination coefficient.

We add the abundance of doubly ionised helium $y^{++}$ to calculate $Y$
when He II $\lambda$4686 is detected. We take the oxygen abundance from
IT04. The final results are displayed for the mass fraction of helium $Y$
as a function of the oxygen abundance relative to hydrogen O/H, and
linear extrapolation is employed to obtain the primordial helium abundance 
$Y_p$ in the zero oxygen abundance limit (Peimbert \& Torres-Peimbert 1974),
\begin{equation}
Y=Y_p+\frac{dY}{d({\rm O/H})}\left(\frac{\rm O}{\rm H}\right).
\end{equation}
We adopt $dY/d({\rm O/H})=18.2 dY/dZ$ from IT04.

\section{Results}

Extinction and stellar absorption of the hydrogen lines are determined from
the 4 Balmer lines. The extinction parameter $c_{\rm H\beta}$ agrees very 
well with those of IT04 in spite of the different extinction curves used.
Curiously, the stellar absorption we derived is generally somewhat
larger than that of IT04, although we do not claim that
the two are inconsistent since errors are generally large. 
For most of the case absorption strengths lie in the range of $0-6$\AA,
but for two it is larger than 10\AA. The mean is 
$a_{\rm HI}=2.5\pm3.5$\AA, where the error stands for the standard
deviation. This is a natural value for HII regions.
As noted by OS04, we find negative values for
absorptions for 11 HII regions. This occurrence is expected from a large
scatter in $a_{\rm HI}$, but might be ascribed to 
the collisional excitation for hydrogen emission lines that are not taken
into account. In any case, the error induced by resetting the negative
absorption width to zero will be small for the helium emissivity analysis,
since the reference H$\beta$ has a large equivalent width $>200$\AA.

To confirm that our results agree with those of IT04 when He I stellar
absorption is
ignored, we first carry out the analysis assuming  $a_{\rm HeI}=0$. 
The helium abundance plotted as a function of the oxygen abundance 
[O/H] presented in Figure 1(a) confirms the trend seen in IT04,
although individual data scatter more with our analysis. 
We obtain $Y_p=0.234\pm0.003$ consistent with $0.2385\pm 0.0015$
 from the extrapolation  of the results tabulated in IT04 [see Figure
2(a)]\footnote{
IT04 give $Y$ for individual HII regions but do not give 
$Y_p$ for their new data alone. 
They combined with their 1998 data and present 
$Y_p = 0.2429\pm 0.0009$. } 
Our $\chi^2$ is rather poor: the mean is $\overline{\chi^2}=5.9$. 
There are five cases
(HS 0122+0743, Mrk724, POX36, HS0128+2832 and Mrk 1236) that give 
$\chi^2>10$, implying the inadequacy of the procedure we assumed.      

When we include $a_{\rm HeI}$ as a free parameter, and carry out
a similar analysis, $\chi^2$ is enormously improved: we find
$\overline{\chi^2}=2.0$, which is dominated by a few bad fits,
notably by one system giving $\chi^2=6.1$ with the zero He I absorption width.
For all cases, except one, that showed poor $\chi^2$ 
with $a_{\rm HeI}=0$, fits are greatly improved. In particular
for the five cases of poor fit we noted above  
$\chi^2$ drops from 29.5 to 1.2 for HS0122+0743, 10.1 to 2.5
for Mrk724, 10.1 to 1.0 for POX36. and 18.5 to 0.9
for HS0128+2832, and from 10.6 to 4.4 for
Mrk 1236, indicating the need to take the stellar absorption 
into account. 

Figure 1(b) shows the plot that can be compared with Figure 1(a)
but with a non-vanishing $a_{\rm HeI}$. It is interesting to
see that the derived helium abundance is more strongly
affected for metal poor HII regions, pushing up the
helium abundance extrapolated to the zero metallicity quite
a bit and resulting in $Y_p=0.250\pm 0.004$. 
Note that the large scatter around the extrapolation line seen
in panel (a) is not visible in this plot any more.
The $\chi^2$ curves are given in Figure 2(a) for $Y_p$ and 
for $dY/d({\rm O/H})$ (b) which is discussed below.

The absorption widths derived from our fit have generally large
errors, $\approx 50$\%. 
The mean absorption equivalent widths are $a_{\rm HeI}=0.40\pm0.31$.
There are 6 HII regions for which the central values of $a_{\rm HeI}$
are negative. 
The four among six are the cases for which 
the fits without stellar absorption gave unusually good $\chi^2$.
The only cases that are marginally acceptable
are with HS1028+3843 ($\chi^2=3.6$) and Mrk35 ($\chi^2=6.1$).
(In the analysis with $a_{\rm HeI}=0$, 10 systems have $\chi^2$
greater than 6.1.)

OS04 presented their analysis for seven HII regions common in our sample.
A comparison shows that our Y is always consistent with theirs within
one standard deviation, although the errors are significantly larger
in OS04 due to larger uncertainties in the temperature of the plasma.
We find other parameters, such as temperature, are also consistent.

Figure 3 shows the average absorption equivalent width that varies with
the heavy element abundance. Although individual data show a rather large
scatter, the trend is clear in this binned plot: the stellar absorption
effect is more important for metal poor HII regions\footnote{While
our derived parameters are consistent with those of OS04, the
effect is not discernible in OS04
due to larger errors (introduced by the temperature uncertainty)
and their smaller data set.}. 
Since the
oxygen abundance shows a tight anticorrelation with temperature of
the plasma, this figure is also interpreted as showing
the correlation of the stellar absorption with the plasma temperature:
Higher the temperature, more the important absorption.
This is the systematic trend that largely modifies the extrapolation of the 
helium abundance to the zero metallicity. 

In this connection another interesting quantity is $dY/d({\rm O/H)}$,
i.e., the increment of helium per heavy element production. The IT04
results give $dY/d({\rm O/H})=82\pm 15$ ( or 
$dY/dZ=4.5\pm0.8$, which is consistent with the final result
IT04 quoted, $dY/dZ=3.7 \pm 1.2$) from a sample of seven HII regions. 
Our analysis with $a_{\rm HeI}$
set equal to zero gives $dY/d({\rm O/H})=86 \pm 18$ or
$dY/dZ=4.7 \pm 1.0$.  On the other hand, with stellar absorption
we obtain $dY/d({\rm O/H})=20 \pm 25$ or $dY/dZ=1.1 \pm 1.4$, which is a
drastic decrease.

We remark that IT04 use only 3 lines in deriving the helium 
abundance, dropping $\lambda 3889$, 
$\lambda 4026$ and $\lambda7065$, with the anticipation that
the other 3 bright lines are less affected by stellar absorption,
while they use $\lambda 3889$ and $\lambda7065$ to constrain
other parameters. We also tried to drop the three lines when we
calculate $Y$, but the results remain unchanged from the full 
6 line analysis.

\section{Conclusions}

Using the IT04 sample we showed that the neglect of the stellar absorption 
on the helium emission lines causes a large systematic effect 
on $Y_p$ and $dY/dZ$.  This is due to the fact that the stellar
absorption equivalent width shows a trend that 
increases towards metal poor systems.
The inclusion of helium stellar absorption improves enormously
the acceptance of fit to the He I emission data, especially when
the quality was bad without the inclusion of stellar absorption. 
The resulting magnitude of the absorption equivalent width of
He I $\lambda4471$ line (and also of the H I Balmer lines) is on 
the order that is expected in a population synthesis calculation  
(Gonz\'alez Delgado et al. 1999) for the nebula phase. 

With the inclusion of stellar absorption we obtained the primordial
helium abundance increased from $y_p=0.234\pm 0.004$ to
$0.250\pm 0.004$.  We do not claim that the latter is the true value,
but it is much preferred to the former on the ground of
much smaller $\chi^2$ for the fit to individual HII regions,
while the 6-line fit without stellar absorption is barely
acceptable.  Or, most conservatively, one can claim that we 
cannot obtain the primordial helium abundance to the error 
of $\delta Y_p\approx 0.004$ or less unless underlying
stellar absorption is properly understood.

In our analysis we noted that the minimisation of the sum of
the $\chi^2$ over individual lines, ignoring the correlation among
lines, is likely to underestimate the error.

If we accept our higher helium abundance, we find $n_B/n_\gamma=
7.9^{+4.0}_{-2.4}\times 10^{-10}$ with the aid of the standard Big Bang nucleosynthesis
calculation (e.g., Olive et al. 2001). 
This baryon abundance is consistent with that inferred from
cosmic microwave background anisotropies (Spergel et al. 2003).

We also found that $dY/dZ$ is largely affected upon the inclusion of stellar
absorption:   the original value of $dY/dZ\approx 4-5$ decreases to
$1\pm 1$. This smaller number is consistent with the derivative inferred
form the standard solar model (Bahcall et al. 2001) 
$\Delta Y/\Delta Z=(Y_{\rm initial}-Y_p)/Z_{\rm initial}=1.4$,
and also with $2.1\pm 0.4$ from the 
dwarf star atmosphere (Jimenez et al. 2003) using the method 
introduced by Pagel and Portinari (1998) who presented $3\pm2$
for this derivative.

\acknowledgments 

This work is supported in part by Grants in Aid of the Ministry of 
Education of Japan at Kashiwa. MF received support from the Monell Foundation
at Princeton.

\newpage
 

\begin{deluxetable}{lccccccccccc}
\tabletypesize{\scriptsize}
\rotate
\tablecaption{Physical Properties of HII Regions and Helium Abundance \label{tab:HII}}
\tablewidth{0pt}
\tablehead{
\colhead{HII reion} & 
\colhead{O/H [$\times 10^{-4}$]} & 
\colhead{$T_{e}[10^4 K]$} & 
\colhead{$C_{{\rm H}_\beta}$} & 
\colhead{$a({\rm HI})$ [\AA]}  &
\colhead{$n_{e}$ [cm$^{-3}$]} & 
\colhead{$\tau$} &
\colhead{$a({\rm HeI})$ [\AA]} & 
\colhead{$Y$} & \colhead{$\chi^2$} &
\colhead{$Y$ (w/o abs)} & 
\colhead{$\chi^2$ (w/o abs)} 
}
\startdata
   J 0519+0007 & 
     $0.270 \pm 0.009$ & 
     $2.073\pm 0.034 $ &
     $0.254\pm 0.020 $ & $0.00 \pm 0.27$ &
     $210^{~+159}_{~-139}$ & $3.76^{~+0.75}_{~-0.78}$ & 
     $0.49^{~+0.41}_{~-0.34}$ &
     $0.2586^{~+0.0134}_{~-0.0145}$ & $0.77$ & 
     $0.2448^{~+0.0115}_{~-0.0111}$ & $2.98$\\[0.4em]
   HS 2236+1344 & 
     $0.296 \pm 0.008$ & 
     $2.122\pm 0.029 $ &
     $0.134\pm 0.024 $ & $4.98 \pm 1.31$ &
     $170^{~+157}_{~-142}$ & $4.56^{~+1.06}_{~-0.99}$ & 
     $0.17^{~+0.46}_{~-0.17}$ &
     $0.2458^{~+0.0139}_{~-0.0132}$ & $2.99$ & 
     $0.2410^{~+0.0093}_{~-0.0091}$ & $3.13$\\[0.4em]
   HS 0122+0743 & 
     $0.397 \pm 0.011$ & 
     $ 1.786\pm 0.022 $ &
     $0.121\pm 0.024 $ & $3.75 \pm 1.19$ &
     $0^{~+123}_{~}$ & $0.92^{~+0.41}_{~-0.78}$ & 
     $1.20^{~+0.35}_{~-0.29}$ &
     $0.2612^{~+0.0049}_{~-0.0091}$ & $1.21$ & 
     $0.2277^{~+0.0067}_{~-0.0071}$ & $29.5$\\[0.4em]
   HS 0837+4717 & 
     $0.398 \pm 0.010$ & 
     $ 1.952\pm 0.023 $ &
     $0.246\pm 0.019 $ & $0.00 \pm 0.09$ &
     $305^{~+88}_{~-101}$ & $6.04^{~+0.52}_{~-0.53}$ & 
     $0.00^{~+0.24}_{~}$ &
     $0.2541^{~+0.0088}_{~-0.0064}$ & $0.88$ & 
     $0.2541^{~+0.0068}_{~-0.0064}$ & $0.88$\\[0.4em]
   CGCG007-025(2) & 
     $0.547 \pm 0.017$ & 
     $ 1.665\pm 0.026 $ &
     $0.202\pm 0.019 $ & $0.00 \pm 0.21$ &
     $0^{~+43}_{~}$ & $1.36^{~+0.49}_{~-0.45}$ & 
     $0.47^{~+0.39}_{~-0.32}$ &
     $0.2508^{~+0.0056}_{~-0.0055}$ & $4.73$ & 
     $0.2448^{~+0.0035}_{~-0.0041}$ & $7.09$\\[0.4em]
   CGCG007-025(1) & 
     $0.596 \pm 0.014$ & 
     $ 1.651\pm 0.017 $ &
     $0.278\pm 0.018 $ & $0.00 \pm 0.10$ &
     $115^{~+116}_{~-102}$ & $1.44^{~+0.50}_{~-0.50}$ & 
     $0.40^{~+0.24}_{~-0.22}$ &
     $0.2568^{~+0.0071}_{~-0.0075}$ & $2.93$ & 
     $0.2466^{~+0.0051}_{~-0.0054}$ & $6.38$\\[0.4em]
   HS 0134+3415 & 
     $0.720 \pm 0.019$ & 
     $1.643\pm 0.017 $ &
     $0.079\pm 0.018 $ & $0.00 \pm 0.24$ &
     $0^{~+163}_{~}$ & $1.92^{~+0.39}_{~-0.72}$ & 
     $0.42^{~+0.40}_{~-0.36}$ &
     $0.2566^{~+0.0049}_{~-0.0111}$ & $0.05$ & 
     $0.2478^{~+0.0068}_{~-0.0091}$ & $1.44$\\[0.4em]
   HS 1028+3843 & 
     $0.781 \pm 0.020$ & 
     $1.593\pm 0.016 $ &
     $0.009\pm 0.023 $ & $2.07 \pm 1.83$ &
     $465^{~+158}_{~-161}$ & $5.48^{~+1.01}_{~-0.94}$ & 
     $0.0^{~+0.16}_{~}$ &
     $0.2571^{~+0.0072}_{~-0.0072}$ & $3.60$ & 
     $0.2573^{~+0.0070}_{~-0.0074}$ & $3.60$\\[0.4em]
   HS 0811+4913 & 
     $0.928 \pm 0.024$ & 
     $1.451\pm 0.015 $ &
     $0.116\pm 0.024 $ & $6.21 \pm 1.80$ &
     $195^{~+234}_{~-195}$ & $0.32^{~+1.05}_{~-0.32}$ & 
     $0.67^{~+0.55}_{~-0.48}$ &
     $0.2437^{~+0.0114}_{~-0.0113}$ & $1.03$ & 
     $0.2325^{~+0.0066}_{~-0.0077}$ & $3.22$\\[0.4em]
   HS 1214+3801 & 
     $1.044 \pm 0.024$ & 
     $1.344\pm 0.012 $ &
     $0.329\pm 0.024 $ & $4.59 \pm 1.16$ &
     $285^{~+110}_{~-206}$ & $0.00^{~+0.71}_{~}$ & 
     $0.45^{~+0.23}_{~-0.19}$ &
     $0.2463^{~+0.0073}_{~-0.0054}$ & $2.67$ & 
     $0.2374^{~+0.0041}_{~-0.0050}$ & $9.04$\\[0.4em]
   Mrk 724 & 
     $1.076 \pm 0.029$ & 
     $1.298\pm 0.014 $ &
     $0.122\pm 0.018 $ & $0.00 \pm 0.12$ &
     $80^{~+152}_{~-80}$ & $0.20^{~+0.58}_{~-0.20}$ & 
     $0.19^{~+0.09}_{~-0.08}$ &
     $0.2524^{~+0.0056}_{~-0.0062}$ & $2.46$ & 
     $0.2429^{~+0.0040}_{~-0.0044}$ & $10.1$\\[0.4em]
   HS 0029+1748 & 
     $1.101 \pm 0.035$ & 
     $1.289\pm 0.015 $ &
     $0.385\pm 0.026 $ & $5.11 \pm 1.01$ &
     $0^{~+174}_{~}$ & $1.60^{~+0.48}_{~-0.71}$ & 
     $0.27^{~+0.27}_{~-0.22}$ &
     $0.2468^{~+0.0057}_{~-0.0064}$ & $2.58$ & 
     $0.2420^{~+0.0034}_{~-0.0075}$ & $4.11$\\[0.4em]
   Mrk 67 & 
     $1.116 \pm 0.046$ & 
     $1.320\pm 0.023 $ &
     $0.190\pm 0.026 $ & $1.99 \pm 0.58$ &
     $135^{~+628}_{~-135}$ & $1.76^{~+1.08}_{~-1.42}$ & 
     $0.35^{~+0.23}_{~-0.22}$ &
     $0.2607^{~+0.0116}_{~-0.0223}$ & $0.40$ & 
     $0.2270^{~+0.0147}_{~-0.0039}$ & $2.84$\\[0.4em]
   POX 36 & 
     $1.131 \pm 0.056$ & 
     $1.256\pm 0.029 $ &
     $0.150\pm 0.019 $ & $0.00 \pm 0.08$ &
     $5^{~+147}_{~-5}$ & $0.00^{~+0.46}_{~}$ & 
     $0.40^{~+0.18}_{~-0.16}$ &
     $0.2551^{~+0.0054}_{~-0.0065}$ & $0.95$ & 
     $0.2452^{~+0.0034}_{~-0.0055}$ & $10.1$\\[0.4em]
   UM 439 & 
     $1.167 \pm 0.029$ & 
     $1.415\pm 0.013 $ &
     $0.234\pm 0.024 $ & $1.45 \pm 0.78$ &
     $205^{~+232}_{~-205}$ & $3.52^{~+0.96}_{~-0.91}$ & 
     $0.00^{~+0.11}_{~}$ &
     $0.2446^{~+0.0080}_{~-0.0079}$ & $0.92$ & 
     $0.2447^{~+0.0077}_{~-0.0080}$ & $0.92$\\[0.4em]
   HS 0924+3821 & 
     $1.178 \pm 0.045$ & 
     $1.261\pm 0.020 $ &
     $0.159\pm 0.025 $ & $2.59 \pm 0.66$ &
     $0^{~+306}_{~}$ & $0.88^{~+0.78}_{~-0.88}$ & 
     $0.50^{~+0.24}_{~-0.20}$ &
     $0.2557^{~+0.0060}_{~-0.0095}$ & $0.38$ & 
     $0.2222^{~+0.0146}_{~-0.0035}$ & $7.12$\\[0.4em]
   Mrk 450 (2) & 
     $1.195 \pm 0.064$ & 
     $1.247\pm 0.028 $ &
     $0.058\pm 0.027 $ & $4.16 \pm 1.20$ &
     $675^{~+\cdot\cdot\cdot}_{~-562}$ & $0.32^{~+1.25}_{~-0.32}$ & 
     $0.49^{~+0.65}_{~-0.49}$ &
     $0.2436^{~+0.0178}_{~-0.0156}$ & $2.21$ & 
     $0.2308^{~+0.0154}_{~-0.0054}$ & $3.00$\\[0.4em]
   UM 422 & 
     $1.291 \pm 0.036$ & 
     $1.300\pm 0.013 $ &
     $0.119\pm 0.021 $ & $0.00 \pm 0.97$ &
     $95^{~+293}_{~-95}$ & $0.56^{~+0.73}_{~-0.56}$ & 
     $0.58^{~+0.38}_{~-0.35}$ &
     $0.2581^{~+0.0076}_{~-0.0107}$ & $2.08$ & 
     $0.2458^{~+0.0068}_{~-0.0064}$ & $5.08$\\[0.4em]
   HS 0128+2832 & 
     $1.370 \pm 0.033$ & 
     $1.261\pm 0.010 $ &
     $0.408\pm 0.028 $  & $11.6 \pm 1.8$ &
     $285^{~+386}_{~-285}$ & $2.72^{~+1.29}_{~-1.14}$ & 
     $1.10^{~+0.38}_{~-0.32}$ &
     $0.2561^{~+0.0107}_{~-0.0108}$ & $0.93$ & 
     $0.2271^{~+0.0031}_{~-0.0027}$ & $18.5$\\[0.4em]
   Mrk 1236 & 
     $1.403 \pm 0.036$ & 
     $1.229\pm 0.012 $ &
     $0.356\pm 0.024 $ & $0.04 \pm 1.13$ &
     $0^{~+103}_{~}$ & $2.00^{~+0.31}_{~-0.45}$ & 
     $0.67^{~+0.33}_{~-0.29}$ &
     $0.2588^{~+0.0048}_{~-0.0045}$ & $4.36$ & 
     $0.2500^{~+0.0029}_{~-0.0053}$ & $10.6$\\[0.4em]
   NGC 346 (A) & 
     $1.416 \pm 0.208$ & 
     $1.313\pm 0.009 $ &
     $0.163\pm 0.009 $ & $0.87 \pm 0.48$ &
     $0^{~+52}_{~}$ & $0.16^{~+0.09}_{~-0.16}$ & 
     $0.15^{~+0.13}_{~-0.11}$ &
     $0.2485^{~+0.0021}_{~-0.0026}$ & $2.33$ & 
     $0.2462^{~+0.0013}_{~-0.0018}$ & $4.04$\\[0.4em]
   Mrk 450 (1) & 
     $1.422 \pm 0.039$ & 
     $1.170\pm 0.012 $ &
     $0.144\pm 0.023 $ & $2.96 \pm 1.12$ &
     $170^{~+378}_{~-170}$ & $2.96^{~+0.90}_{~-1.16}$ & 
     $0.60^{~+0.30}_{~-0.26}$ &
     $0.2537^{~+0.0066}_{~-0.0075}$ & $1.79$ & 
     $0.2398^{~+0.0062}_{~-0.0065}$ & $7.61$\\[0.4em]
   UM 238 & 
     $1.462 \pm 0.049$ & 
     $1.250\pm 0.015 $ &
     $0.235\pm 0.030 $ & $15.5 \pm 2.9$ &
     $700^{~+\cdot\cdot\cdot}_{~-481}$ & $2.60^{~+1.69}_{~-0.97}$ & 
     $0.98^{~+0.71}_{~-0.65}$ &
     $0.2473^{~+0.0145}_{~-0.0131}$ & $1.29$ & 
     $0.2324^{~+0.0064}_{~-0.0039}$ & $4.03$\\[0.4em]
   HS 0735+3512 & 
     $1.493 \pm 0.043$ & 
     $1.206\pm 0.014 $ &
     $0.236\pm 0.023 $ & $0.16 \pm 0.61$ &
     $180^{~+447}_{~-180}$ & $2.44^{~+1.02}_{~-1.20}$ & 
     $0.25^{~+0.22}_{~-0.20}$ &
     $0.2576^{~+0.0083}_{~-0.0108}$ & $1.44$ & 
     $0.2489^{~+0.0079}_{~-0.0099}$ & $2.98$\\[0.4em]
   HS 2359+1659 & 
     $1.524 \pm 0.056$ & 
     $1.192\pm 0.016 $ &
     $0.324\pm 0.025 $ & $1.33 \pm 1.16$ &
     $145^{~+497}_{~-145}$ & $1.56^{~+0.99}_{~-1.21}$ & 
     $0.00^{~+0.25}_{~}$ &
     $0.2470^{~+0.0065}_{~-0.0091}$ & $0.74$ & 
     $0.2470^{~+0.0060}_{~-0.0091}$ & $0.74$\\[0.4em]
   HS 1311+3628 & 
     $1.525 \pm 0.045$ & 
     $1.141\pm 0.013 $ &
     $0.066\pm 0.024 $ & $1.48 \pm 1.30$ &
     $450^{~+\cdot\cdot\cdot}_{~-450}$ & $0.56^{~+1.30}_{~-0.56}$ & 
     $0.63^{~+0.46}_{~-0.44}$ &
     $0.2527^{~+0.0085}_{~-0.0114}$ & $1.58$ & 
     $0.2401^{~+0.0064}_{~-0.0028}$ & $3.78$\\[0.4em]
   HS 1213+3636 & 
     $1.561 \pm 0.087$ & 
     $1.077\pm 0.027 $ &
     $0.019\pm 0.019 $ & $0.00 \pm 0.61$ &
     $50^{~+193}_{~-50}$ & $0.00^{~+0.44}_{~}$ & 
     $0.48^{~+0.36}_{~-0.32}$ &
     $0.2597^{~+0.0056}_{~-0.0057}$ & $3.50$ & 
     $0.2533^{~+0.0030}_{~-0.0038}$ & $5.91$\\[0.4em]
   UM 396 & 
     $1.708 \pm 0.064$ & 
     $1.140\pm 0.015 $ &
     $0.258\pm 0.028 $ & $6.31 \pm 1.22$ &
     $620^{~+\cdot\cdot\cdot}_{~620}$ & $1.20^{~+1.71}_{~-0.75}$ & 
     $0.00^{~+0.21}_{~}$ &
     $0.2471^{~+0.0094}_{~-0.0078}$ & $1.51$ & 
     $0.2471^{~+0.0081}_{~-0.0078}$ & $1.52$\\[0.4em]
   Mrk 1315 & 
     $1.774 \pm 0.041$ & 
     $1.101\pm 0.008 $ &
     $0.149\pm 0.017 $ & $0.00 \pm 0.31$ &
     $310^{~+249}_{~-231}$ & $0.32^{~+0.66}_{~-0.32}$ & 
     $0.15^{~+0.24}_{~-0.15}$ &
     $0.2554^{~+0.0041}_{~-0.0043}$ & $2.85$ & 
     $0.2533^{~+0.0031}_{~-0.0030}$ & $3.34$\\[0.4em]
   Mrk 1329 & 
     $1.787 \pm 0.044$ & 
     $1.079\pm 0.009 $ &
     $0.178\pm 0.023 $ & $0.82 \pm 0.97$ &
     $500^{~+\cdot\cdot\cdot}_{~-460}$ & $0.48^{~+1.24}_{~-0.48}$ & 
     $0.31^{~+0.22}_{~-0.23}$ &
     $0.2574^{~+0.0055}_{~-0.0073}$ & $1.71$ & 
     $0.2497^{~+0.0029}_{~-0.0023}$ & $3.90$\\[0.4em]
   Mrk 35 & 
     $1.977 \pm 0.058$ & 
     $1.016\pm 0.012 $ &
     $0.230\pm 0.018 $ & $0.00 \pm 0.07$ &
     $335^{~+270}_{~-237}$ & $2.24^{~+0.72}_{~-0.71}$ & 
     $0.00^{~+0.08}_{~}$ &
     $0.2554^{~+0.0027}_{~-0.0023}$ & $6.10$ & 
     $0.2555^{~+0.0021}_{~-0.0024}$ & $6.10$\\[0.4em]

\enddata
\tablecomments{The dots in upward errors of $n_e$ means that one sigma
value is not reached within the limit ($n_e=1000$) of our fit we set.} 
\end{deluxetable}

\begin{figure}
\includegraphics[scale=.70]{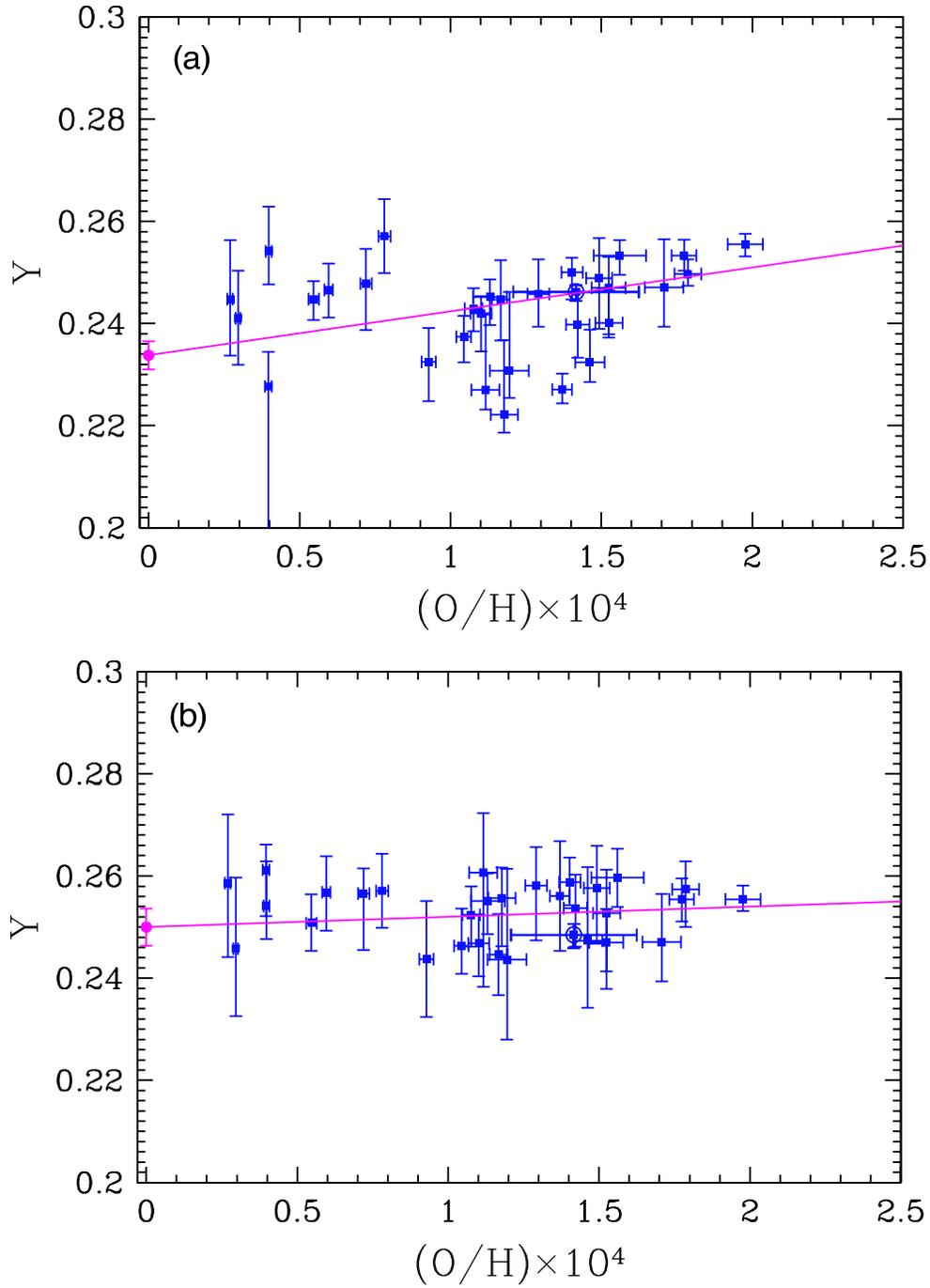}
\caption{Helium mass fraction $Y$ vs oxygen abundance O/H
for 30 HII regions in IT04 (filled circle) and NGC346A (open circle)
(a) without and (b) with steller absorption.
The solid line represents the linear fit.  \label{fig1}}
\end{figure}

\begin{figure}
\includegraphics[scale=.70]{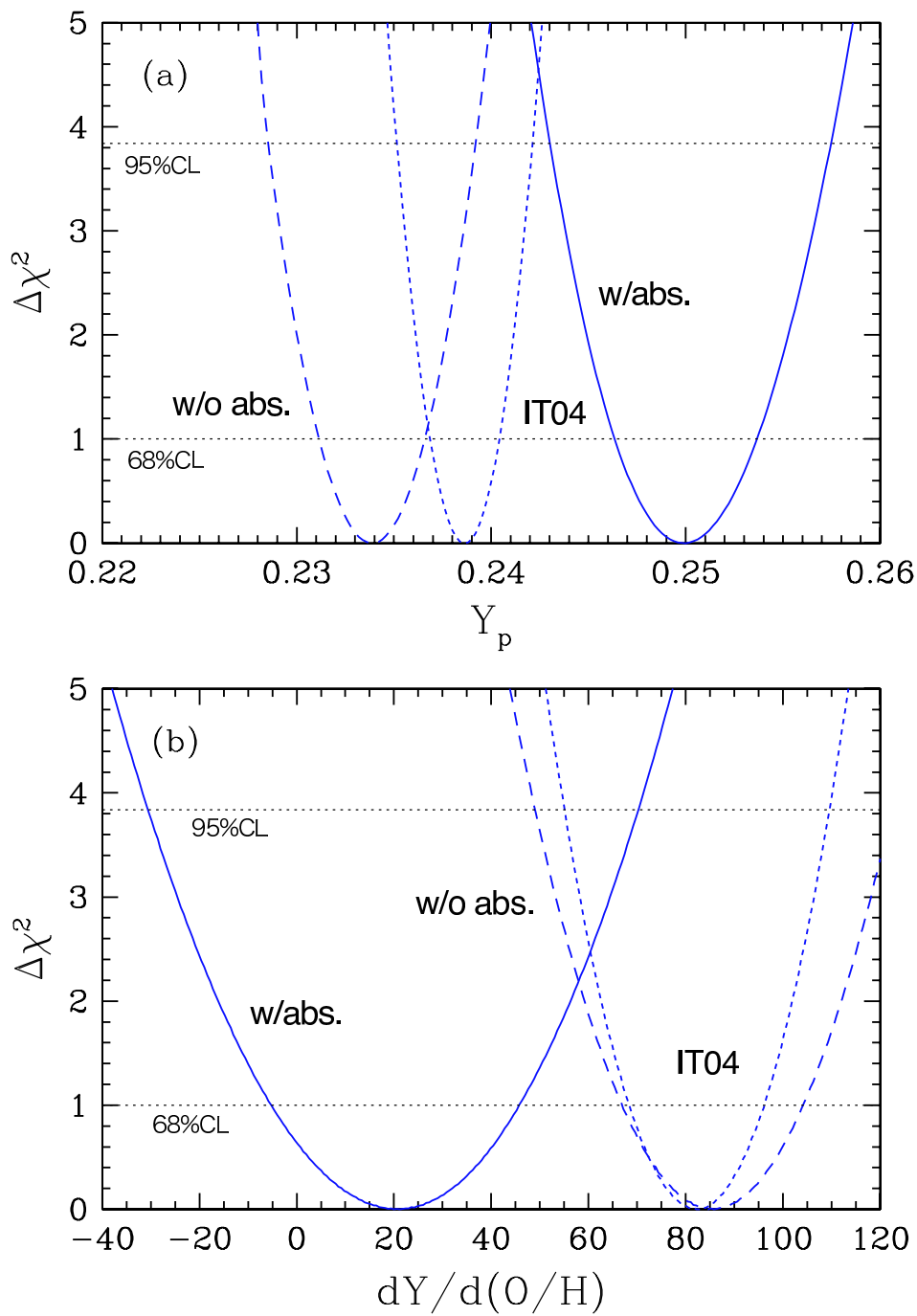}
\caption{$\Delta \chi^2 = \chi^2 - \chi_{\rm min}^2$ 
for (a) $Y_p$ and (b)$dY/d({\rm O/H})$.
The solid and  dashed curves denote the results
with and without absorption, respectively. The dotted curve 
shows the helium abundances given
in IT04. $\Delta\chi^2$ for
68\% and 95\% confidence levels are indicated. \label{fig2}}
\end{figure}

\begin{figure}
\plotone{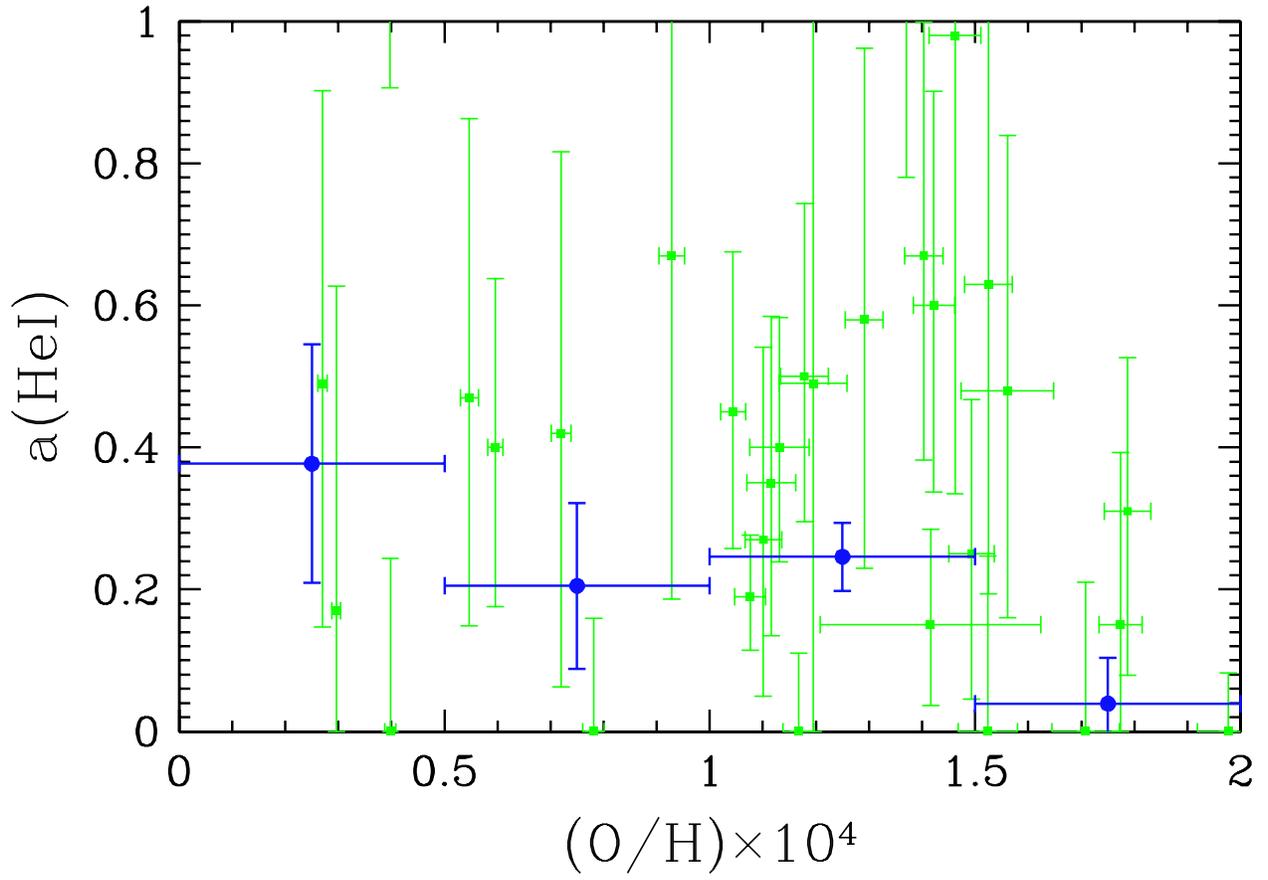}
\caption{Stellar absorption for HeI emission lines 
vs the oxygen abundance (filled square).
The binned average is shown by solid circles.\label{fig3}}
\end{figure}
 

\begin{thebibliography}{}


\bibitem[Bahcall et al.(2001)]{2001ApJ...555..990B} Bahcall, J.~N., 
Pinsonneault, M.~H., \& Basu, S.\ 2001, \apj, 555, 990 

\bibitem[Benjamin et al.(1999)]{1999ApJ...514..307B} Benjamin, R.~A., 
Skillman, E.~D., \& Smits, D.~P.\ 1999, \apj, 514, 307 

\bibitem[Benjamin et al.(2002)]{2002ApJ...569..288B} Benjamin, R.~A., 
Skillman, E.~D., \& Smits, D.~P.\ 2002, \apj, 569, 288 

\bibitem[Eidelman et al. (Particle Data Group) (2004)]{} Eidelman, S.
et al. (Particle data Group) 2004, Phys. Lett. B592, 1


\bibitem[Gonz{\'a}lez Delgado et al.(1999)]{1999ApJS..125..489G} 
Gonz{\'a}lez Delgado, R.~M., Leitherer, C., \& Heckman, T.~M.\ 1999, \apjs, 
125, 489 

\bibitem[Hummer \& Storey(1987)]{1987MNRAS.224..801H} Hummer, D.~G., \& 
Storey, P.~J.\ 1987, \mnras, 224, 801

\bibitem[Izotov \& Thuan(1998)]{1998ApJ...500..188I} Izotov, Y.~I., \& 
Thuan, T.~X.\ 1998, \apj, 500, 188 

\bibitem[Izotov \& Thuan(2004)]{2004ApJ...602..200I} Izotov, Y.~I., \& 
  Thuan, T.~X.\ 2004, \apj, 602, 20 (IT04)

\bibitem[Jimenez et al.(2003)]{2003Sci...299.1552J} Jimenez, R., Flynn, C., 
MacDonald, J., \& Gibson, B.~K.\ 2003, Science, 299, 1552 

\bibitem[Lennon et al.(1993)]{1993A&AS...97..559L} Lennon, D.~J., Dufton, 
P.~L., \& Fitzsimmons, A.\ 1993, \aaps, 97, 559 

 \bibitem[Lyubimkov et al.(2000)]{2000MNRAS.316...19L} Lyubimkov, L.~S., 
Lambert, D.~L., Rachkovskaya, T.~M., Rostopchin, S.~I., Tarasov, A.~E., 
Poklad, D.~B., Larionov, V.~M., \& Larionova, L.~V.\ 2000, \mnras, 316, 19 

\bibitem[O'Donnell(1994)]{1994ApJ...422..158O} O'Donnell, J.~E.\ 1994, 
\apj, 422, 158 

\bibitem[Olive \& Skillman(2001)]{2001NewA....6..119O} Olive, K.~A., \& 
Skillman, E.~D.\ 2001, New Astronomy, 6, 119 

\bibitem[Olive \& Skillman(2004)]{2004ApJ...617...29O} Olive, K.~A., \& 
Skillman, E.~D.\ 2004, \apj, 617, 29 (OS04)

\bibitem[Olive et al.(1997)]{1997ApJ...483..788} 
Olive, K.~A., 
Skillman, E.~D.\ 2004,\& Steigman, G.,\ 1997, \apj, 483, 788 


\bibitem[Olive et al.(2000)]{2000PhR...333..389O} Olive, K.~A., Steigman, 
G., \& Walker, T.~P.\ 2000, \physrep, 333, 389 

\bibitem[Pagel \& Portinari(1998)]{1998MNRAS.298..747P} Pagel, B.~E.~J., \& 
Portinari, L.\ 1998, \mnras, 298, 747 

\bibitem[Pagel et al.(1992)]{1992MNRAS.255..325P} Pagel, B.~E.~J., 
Simonson, E.~A., Terlevich, R.~J., \& Edmunds, M.~G.\ 1992, \mnras, 255, 
325 

\bibitem[Peimbert \& Torres-Peimbert(1974)]{1974ApJ...193..327P} Peimbert, 
M., \& Torres-Peimbert, S.\ 1974, \apj, 193, 327 

\bibitem[Peimbert et al.(2000)]{2000ApJ...541..688P} Peimbert, M., 
Peimbert, A., \& Ruiz, M.~T.\ 2000, \apj, 541, 688 

\bibitem[Peimbert et al.(2002)]{2002ApJ...565..668P} Peimbert, A., 
Peimbert, M., \& Luridiana, V.\ 2002, \apj, 565, 668 

\bibitem[Spergel et al.(2003)]{2003ApJS..148..175S} Spergel, D.~N., et al.\ 
2003, \apjs, 148, 175 



\end{thebibliography}
\end{document}